\documentclass{article}

\usepackage{times,mathptm}
\usepackage{amsmath, amsfonts, bm}
\usepackage{graphicx}
\usepackage{natbib}
\usepackage{authblk}

\usepackage{fancyhdr}

\usepackage{algorithm}
\usepackage[noend]{algpseudocode}

\newlength{\npaperwidth}
\begin{document}

\title{\Large Hydraulic Fracture}

\author[1,3]{Joseph B. Walsh}
\author[2]{Stephen R. Brown}

\affil[1]{\normalsize Box 22, Adamsville, RI 02801 USA \linebreak {}}
\affil[2]{\normalsize Earth Resources Laboratory, Massachusetts
  Institute of Technology, Cambridge, MA 02139 USA \linebreak
  Aprovechar Lab L3C, Montpelier, VT 05602 USA}

\maketitle

\footnotetext[3]{Joseph B. Walsh died on 30 August 2017 at the age of
  86 in Adamsville, R.I. Please read more about Joe Walsh's life here:
  Scholz, C. H., Goldsby, D. L., Bernab\'{e}, Y., and Evans,
  B., (2018), Joseph B. Walsh (1930–2017), Eos, \textbf{99},
  https://doi.org/10.1029/2018EO093999. Published on 6 March 2018.}

\begin{abstract}
  We consider a variation of Griffith's analysis of rupture, one which
  simulates the process of hydrofracturing, where fluid forced into a
  crack raises the fluid pressure until the crack begins to
  grow. Unlike that of Griffith, in this analysis fluid pressure drops
  as a hydrofracture grows. We find that growth of the
  fracture depends on the ratio of the compliances of the fluid and the
  fracture, a non-dimensional parameter called $\alpha_0$ here. The
  pressure needed to initiate a hydrofracture is found to be the same
  as that derived by Griffith. Once a fracture initiates, for
  relatively low values of the model parameter $\alpha_0$ ($\alpha_0
  \leq 0.2$) the fracture advances spontaneously to a new radius which
  depends on the value of $\alpha_0$. For $\alpha_0 \leq 0.2$
  further fluid injection is required to increase the fracture radius
  after spontaneous growth stops. For the case where $\alpha_0 > 0.2$
  each increment of fracture growth requires injection of more
  fluid. For the extreme case where $\alpha_0 = 0$ our results are the
  same as Griffith's, i.e., a fracture once initiated grows without
  limit.

  \vspace*{\fill}
  \centering Copyright \textcopyright\ 2022 Stephen R. Brown
  \vskip 3\baselineskip

\end{abstract}

\newpage
\section{Introduction}
Griffith \citep{griffith:1921,griffith:1924}, in his fundamental study
of fracture, chose to analyze the stress intensity and strain energy
associated with an empty crack loaded by normal stress which originate
at a distance far away. Here we consider a variation of Griffith's
analysis, one which simulates hydrofracturing, where fluid forced into
a crack raises the fluid pressure (and lowers the effective normal
stress on the crack walls) until the conditions are energetically
favorable for growth of the fracture. Mathematical analysis of the two
problems is different because applied stress is held constant or
nearly so in Griffith's analysis whereas in the analysis here fluid
pressure drops as a hydrofracture grows.  The work done by the
constant applied load in Griffith's analysis causes the fracture, once
initiated, to grow without limit. We find that the criterion for
fracture initiation in our analysis is the same as that in
Griffith's. On the other hand, we find that although Griffith's
criterion is necessary for initiation, it is not a sufficient
condition for fracture growth in all cases; i.e., though growth is
possible at Griffith's value of the internal fluid pressure, it might
not necessarily proceed without additional stimulus.

\begin{figure}[b]
\centering
\includegraphics[height=0.4\textwidth]{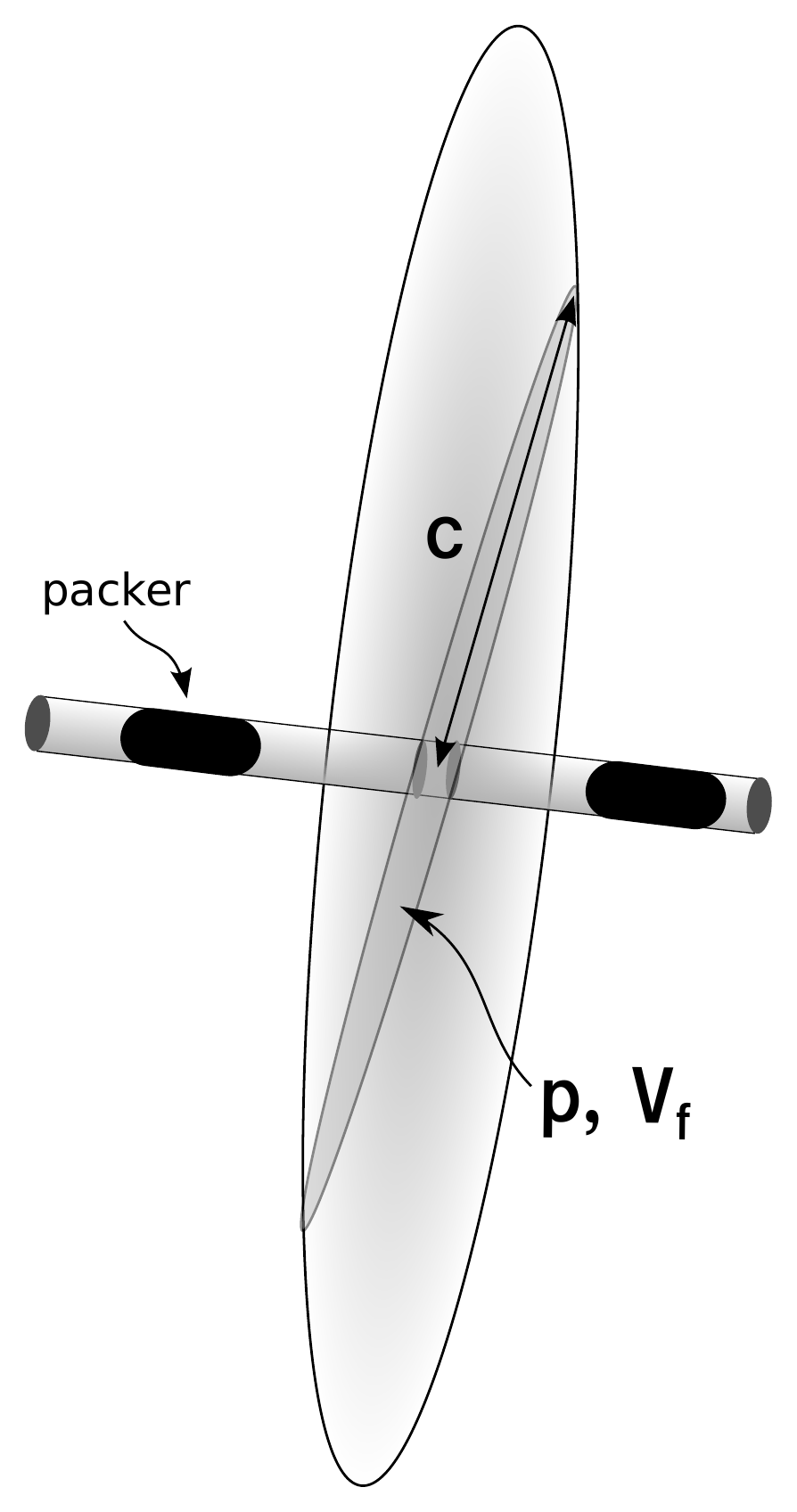}
\caption{Fracture in a borehole. In the analysis we consider the
  cavity to include both the fracture, assumed to be a penny shaped
  crack having radius c, and the drill hole between the packers. Total
  volume is $V_f$. Fluid pressure $p$ is assumed to be uniform
  throughout the cavity and changes in $p$ occur
  instantaneously.\label{fig_1}}
\end{figure}

We model a hydrofracture by the circular crack with radius $c$ seen in
Figure \ref{fig_1}. In the model, before the hydrofracturing
procedure begins, the pressure and temperature of the hydrofracturing
fluid are assumed to be in equilibrium with the surrounding country
rock. The pressures $p$ in the analysis are pressures relative to the
initial equilibrium values and so pressure $p$ is zero when fluid
injection begins.

The country rock surrounding the fracture is assumed to be an
isotropic, elastic, and impermeable infinite medium.  The volume $V_f$
of fluid includes all fluid in contact with the fracture; that is, the
fluid in the fracture and in the drill hole between the packers.  We
don't include time as a parameter and changes in pressure are felt
immediately throughout the fluid volume $V_f$. Time, of course, is an
important parameter and we discuss the effect of it on our results in
a later section.

\section{Analysis}

\subsection{Preliminaries}
As discussed above, for the purposes of analysis we consider the
initial pressure of the system to be zero. To evaluate the effect of
injecting fluid into the cavity we follow Eshelby's
\citep{eshelby:1957} technique for calculating the effect of a
transformational strain upon an inhomogeneity embedded in an infinite
matrix. First, we imagine removing the fluid unit from the fracture
cavity in the country rock and adding a differential volume $dV_i$ to
it. Then, a differential pressure $dp_i$ is applied to the unit to
return the volume to the original cavity size. 
The pressure $dp_i$ needed is:

\begin{equation}
dp_i = \frac{dV_i}{\beta_f V_f},
\label{eqn1}
\end{equation}

\noindent where $\beta_f$ is the compressibility of the fluid and the
volume $V_f$ of fluid is defined above. Once the fluid is compressed
under pressure $dp_i$, it is returned to the cavity, the cavity walls
remaining stress-free.  To equalize stress at the interface, pressure
$dp$ is applied to the both the fluid and to the cavity wall while
contact is maintained between fluid and the surrounding medium. The
decrease $dV_f$ in the volume of the fluid is:

\begin{equation}
dV_f = \beta_f V_f dp,
\label{eqn2}
\end{equation}

\noindent where decrease in volume $dV_f$ is positive. 

The increase $dV_c$ in the volume $V_c$ of the cavity can be found
from (\ref{eqn10}), (\ref{eqn11}), and (\ref{eqn12}) for the
conditions here where fracture radius $c_0$ is constant: i.e.,

\begin{equation}
  dV_c = \frac{16}{9} \frac{\left( 1 - \nu^2 \right)}{\left( 1 - 2\nu \right)} \beta_s c_0^3 dp,
\label{eqn3}
\end{equation}

\noindent
where $\beta_s$ is the compressibility, $\nu$ is the Poisson's ratio
of the elastic matrix. To maintain contact, $dV_f$ must equal $-dV_c$;
and for equilibrium at the interface, pressure $dp_i-dp$ from
(\ref{eqn1}) and (\ref{eqn2}) acting on the fluid must equal pressure
$dp$ from (\ref{eqn3}) acting on the cavity wall. We find from (\ref{eqn1}),
(\ref{eqn2}), and (\ref{eqn3}) with these constraints that the increase $dp$ in pressure
resulting from an injection of the differential fluid volume $dV_i$ is
given by:

\begin{equation}
dp = \left( \frac{dV_i}{\beta_f V_f} \right) \frac{1}{1 + \alpha_0},
\label{eqn4}
\end{equation}

\noindent
where

\begin{equation}
\alpha_0 = \frac{16 (1-\nu^2)}{9 (1-2\nu)} \left( \frac{\beta_s}{\beta_f} \right) \frac{c_0^3}{V_f}.
\label{eqn5}
\end{equation}

\subsection{Griffith Analysis}
Consider the penny-shaped crack in an infinite, impermeable, elastic
medium in Figure \ref{fig_1}. As discussed in the Introduction,
compressible fluid forced into the crack under increasing pressure $p$
via the borehole eventually causes the crack having radius $c_0$ to
grow.  Griffith reasoned that initiation would occur when
conditions were energetically favorable. The components of energy of
interest in this problem are the work $dw_p$ done by the loading
mechanism, here fluid pressure; the change $dw_c$ in the elastic
energy stored in the stressed body; and the energy $dw_s$ needed to
create new surface exposed as the crack advances. Griffith's criterion
for the initiation of growth is simply:

\begin{equation}
dw_s + dw_p + dw_c = 0.
\label{eqn6}
\end{equation}

\noindent
The relationship between surface energy and surface area is generally
taken to be a simple proportionality, i.e.,

\begin{equation}
dw_s = -2 \pi \gamma c dc,
\label{eqn7}
\end{equation}

\noindent where $\gamma$ is the surface energy per unit area. Work
$dw_p$ done by the fluid pressure $p_f$ 
acting on the increase $dV_c$
in the volume of the cavity as the crack advances is

\begin{equation}
dw_p = p dV_c.
\label{eqn8}
\end{equation}

\noindent
Introducing the expression for $dV_c$ from (\ref{eqn3})
  into (\ref{eqn8}), we find:

\begin{equation}
  p dV_c = \frac{16 \left( 1- \nu^2 \right) }{ 9 \left( 1-2 \nu \right)} \beta_s c_0^3 p dp.
\label{eqn9}
\end{equation}

The change $dw_c$ in elastic strain energy is the result of increase
$dc$ in crack radius and decrease $dp$ in fluid pressure; it is
conveniently written in differential form as

\begin{equation}
  dw_c = \frac{\partial w_c}{\partial c} dc + \frac{\partial w_p}{\partial p} dp.
\label{eqn10}
\end{equation}

\noindent The expression for $w_c$ is \citep{sack:1946}:

\begin{equation}
  w_c = \frac{8 (1-\nu^2)}{9 (1-2\nu)} \beta_s c^3 p^2.
\label{eqn11}
\end{equation}

\noindent
The decrease (partial derivative
$\left( \partial w_c / \partial p \right) dp$) in elastic strain
energy in the body arising from the decrease $-dp$ in fluid pressure
is found from (\ref{eqn11})) to be:
\begin{equation}
\left( \frac{\partial w_c}{\partial p} \right) dp = - \frac{16 (1-\nu^2)}{9 (1-2\nu)} \beta_s c_0^3 p dp.
\label{eqn12}
\end{equation}

\noindent Combining (\ref{eqn6})-(\ref{eqn10}), and
(\ref{eqn12}), we find that Griffith's criterion becomes

\begin{equation}
  \frac{\partial w_c}{ \partial c} \geq 2 \pi \gamma c_0.
\label{eqn13}
\end{equation}

Equation (\ref{eqn13}) is Griffith's fracture initiation criterion in
compact form. The partial differential term expresses the increase
in elastic strain energy arising from a differential increase $dc$ in
crack radius carried out with pressure held constant. As established
by Griffith, this must be at least equal to the energy required to
create new surface. Introducing (\ref{eqn11}) into (\ref{eqn13}) and
rearranging gives Griffith's familiar expression for the pressure
$p_G$ at which fracture growth is possible:

\begin{equation}
  p_G = \left[ \frac{3 \pi \left(1-2 \nu \right)}{4 \left( 1- \nu^2 \right)} \left( \frac{\gamma}{\beta_s c_0} \right) \right]^{1/2} .
\label{eqn14}
\end{equation}

\subsection{Fracture Growth}
\citet{griffith:1921,griffith:1924} in his analysis considered the
balance between elastic and dissipative energies for a crack loaded by
stress applied at a remote distance. A consequence of this mode of
loading is that once cracking is initiated, the crack grows without
limit because the initiating stress remains virtually
constant. Hydrofracturing differs not only because the stressed region
is limited to a volume of the order of the crack dimensions, but also
because fluid pressure decreases as the hydrofracture grows.  We now
ask the question: Should we expect a hydraulic fracture to grow
without limit after the fracture is initiated as Griffith cracks do?

We assume that the speed at which the fracture grows, the viscosity of
the hydrofracturing fluid, and any other parameters that involve time
are such that in our analysis fluid always completely fills the cavity
and that the pressure is uniform throughout the fluid. Note that the
mathematical model is the one we have been using in the previous two
sections to analyze pressurizing the system and derive the criterion
for fracture; we now consider behavior as the crack radius grows from
$c_0$ to some arbitrary value $c$.

Let us assume for now that growth occurs without further injection of
fluid after initiation at pressure $p_G$. For fracture growth,
Griffith's criterion must be met at each increment of growth. The rate of change in strain energy $w_c$ when 
the fracture has grown to arbitrary radius $c$ is
found from (\ref{eqn10}) and (\ref{eqn11}) to be:

\begin{equation}
  \left( \frac{\partial w_c}{\partial c} \right) = \frac{8 \left( 1-\nu^2 \right)}{3 \left( 1-2\nu \right)} \beta_s c^2 p^2,
\label{eqn15}
\end{equation}

\noindent
and the rate of change
at initiation, when $c = c_0$, is:

\begin{equation}
  \left( \frac{\partial w_c}{\partial c} \right)_0 = \frac{8 \left( 1-\nu^2 \right)}{3 \left( 1-2\nu \right)} \beta_s c_0^2 p_G^2.
\label{eqn16}
\end{equation}

\noindent
Non-dimensionalizing the criterion by taking the ratio of (\ref{eqn15})
and (\ref{eqn16}), we find:

\begin{equation}
  \left( \frac{\partial w_c}{\partial c} \right) / \left( \frac{\partial w_c}{\partial c} \right)_0 = \left( c / c_0 \right)^2 \left( p / p_G \right)^2.
\label{eqn17}
\end{equation}

To find the pressure $p/p_G$ when the fracture has grown to $c/c_0$,
we first integrate (\ref{eqn4}) at constant fracture radius $c_0$ to find
the injected volume $V_G$ when the fracture is initiated 
(\textit{i.e.}, when $p=p_G$):

\begin{equation}
p_G = \frac{ V_G }{\beta_f V_f} \left( \frac{1}{1 + \alpha_0} \right) .
\label{eqn18}
\end{equation}

\noindent
The pressure $p$ when the fracture radius has increased to $c$, with
$V_G$ remaining constant, is:

\begin{equation}
  p = \frac{V_G}{\beta_f V_f} \left( \frac{1}{1 + \alpha_0 \left( c/c_0 \right)^3} \right).
\label{eqn19}
\end{equation}

Following the procedure leading to (\ref{eqn17}), we find

\begin{equation}
\frac{p}{p_G} =  \frac{1 + \alpha_0}{1 + \alpha_0 \left( c/c_0 \right)^3 },
\label{eqn20}
\end{equation}

\noindent
and (\ref{eqn17}) becomes:

\begin{equation}
  \left( \frac{\partial w_c}{\partial c} \right) / \left( \frac{\partial w_c}{\partial c} \right)_0 = \left[ \left( \frac{c}{c_0} \right) \frac{1 + \alpha_0}{1 + \alpha_0 \left( c/c_0 \right)^3 } \right]^2
\label{eqn21}
\end{equation}

Repeating these steps for the surface energy gives:

\begin{equation}
  \left( \frac{\partial w_s}{\partial c} \right) / \left( \frac{\partial w_s}{\partial c} \right)_0 = \left( \frac{c}{c_0} \right).
\label{eqn22}
\end{equation}

\begin{figure}
\centering
\includegraphics[width=0.6\textwidth]{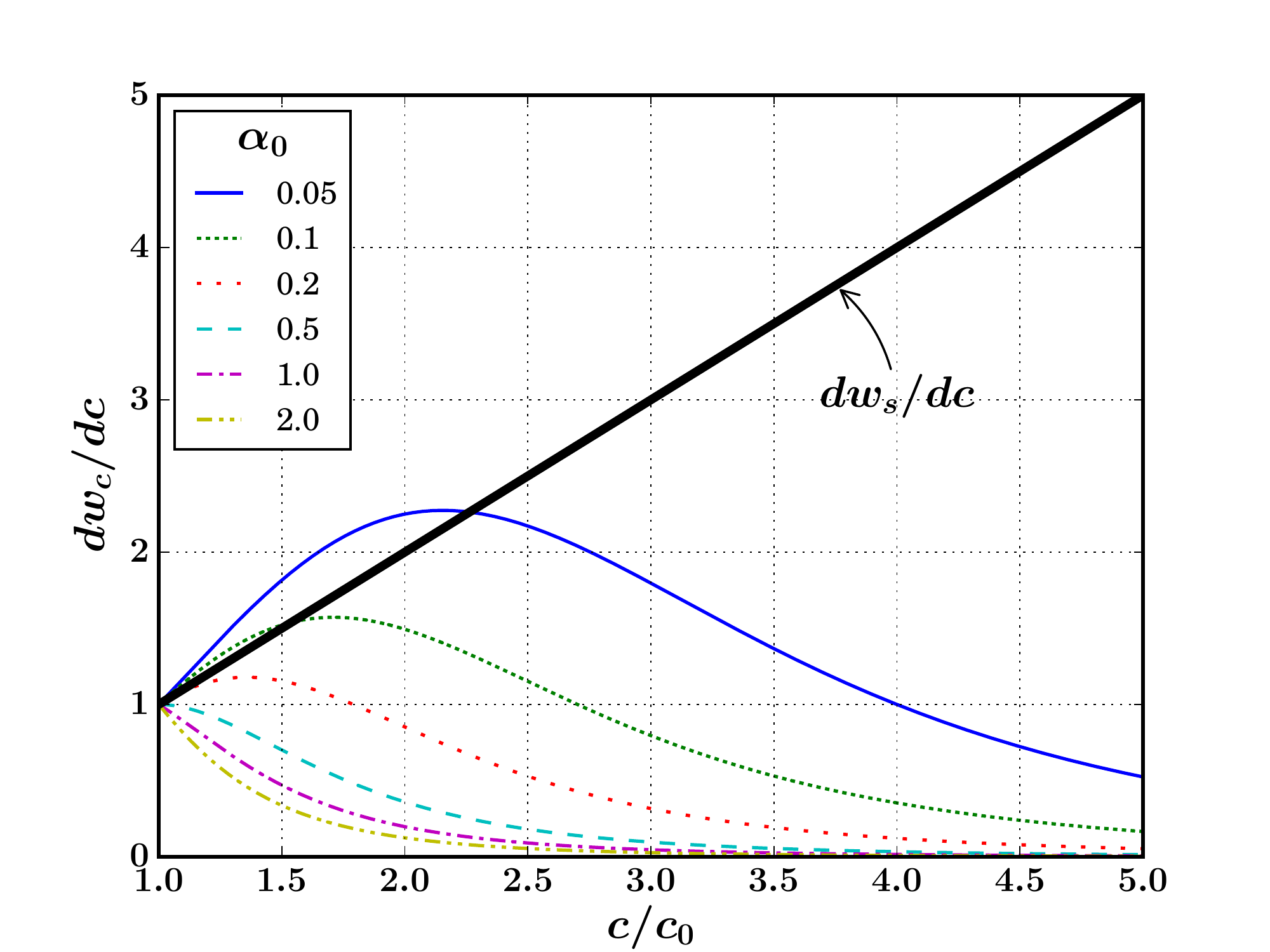}
\caption{The thin lines in the figure are plots of the elastic
  strain energy release rate $dw_c/dc$ and the heavy black line is
  the rate $dw_s/dc$ of fracture energy dissipation; both are plotted
  in the dimensionless form in (\ref{eqn21}) and (\ref{eqn22}) as
  functions of the parameter $\alpha_0$ and crack radius $c/c_0$. For
  fracture growth, the elastic energy release rate must be greater
  than the rate at which fracture energy is dissipated, that is the
  thin lines must be above the heavy black one. We see in the
  figure that growth occurs spontaneously only over a restricted range
  of values of $\alpha_0$; that is, as shown in the text, for
  $\alpha_0 < 0.2$. In cases where $\alpha_0 > 0.2$ and cases where
  $\alpha_0 < 0.2$ but beyond the range of spontaneous growth,
  fracture size can be increased only by injecting fluid.
  Graphically, for any combination of crack length $c/c_0$ and
  parameter $\alpha_0$ where a line falls below the heavy black line,
  the crack is stable and additional internal fluid pressure must be
  increased to make the crack grow. \label{fig_2}}
\end{figure}

The two energy functions in (\ref{eqn21}) and (\ref{eqn22}) are
plotted in Figure \ref{fig_2}. Note in the figure that crack growth is
spontaneous for systems having relatively low values of $\alpha_0$,
but the fracture stops after running a limited distance. Solving
(\ref{eqn21}) and (\ref{eqn22}) shows that spontaneous growth occurs
for $\alpha_0$ in the range:

\begin{equation}
0 \leq \alpha_0 \leq 0.2.
\label{eqn23}
\end{equation}

In the limit as $ \alpha_0 \rightarrow 0$ (i.e., the fracturing fluid
is very compressible as for a gas), we revert to Griffith's model
where the load remains approximately constant and fracture growth is
unlimited. Note in (\ref{eqn20}) that loading pressure $p$ for this
extreme case remains constant at Griffith's value $p_G$.

\begin{figure}
\centering
\includegraphics[width=0.6\textwidth]{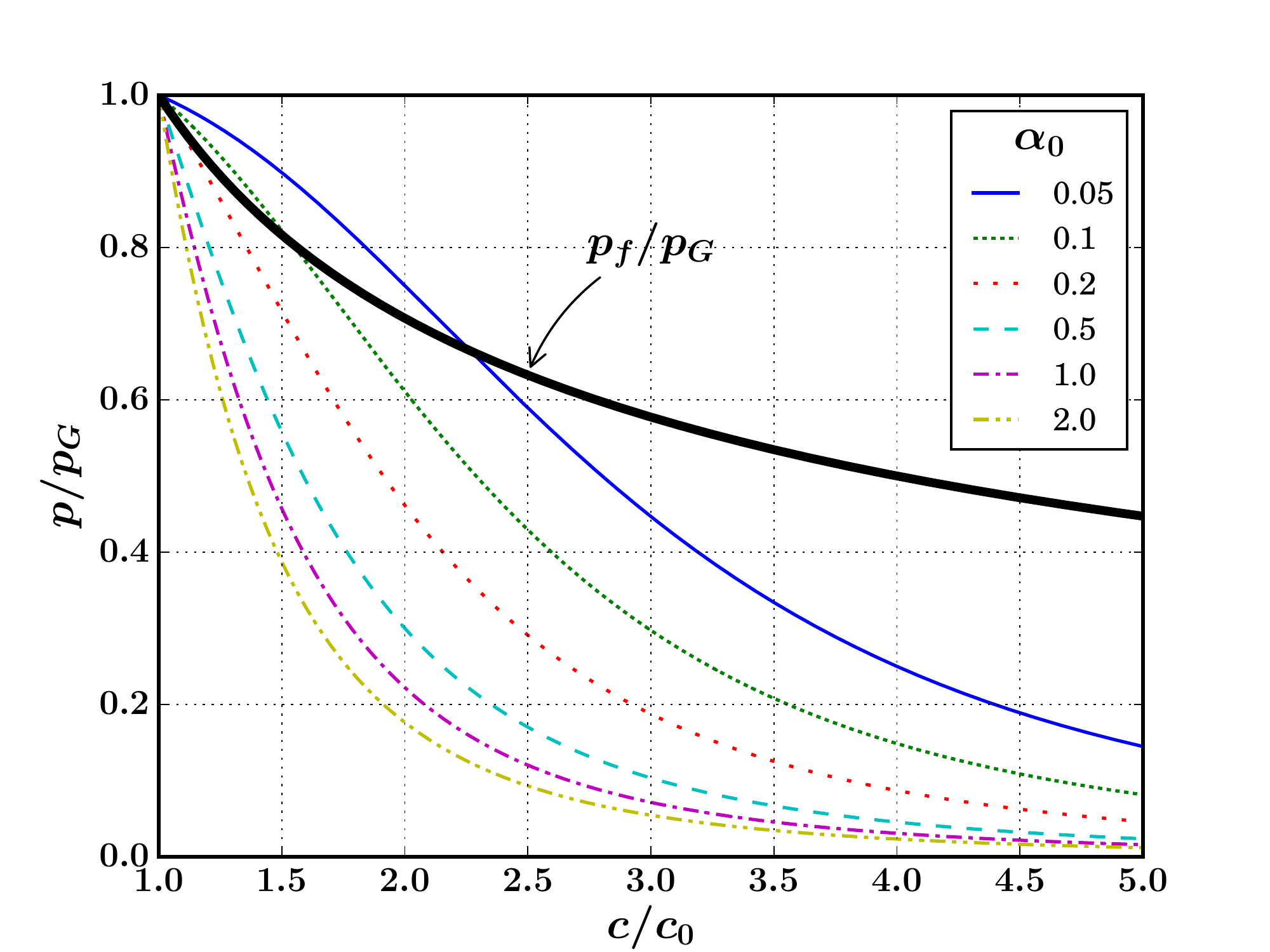}
\caption{The thin lines are plots of non-dimensional fluid
  pressure $p/p_{G}$ after the fracture has grown to some radius
  $c/c_0$ without additional fluid being injected given by (\ref{eqn20}). The
  heavy black line is the non-dimensional fluid pressure $p_f/p_{G}$ needed
  for continued growth as given by (\ref{eqn24}). For any combination of crack
  length $c/c_0$ and parameter $\alpha_0$ where a line falls below
  the heavy black line, the crack is stable and additional internal fluid
  pressure must be increased to make the crack grow. Lines for
  $\alpha_0$ in the range ($0 \leq \alpha_0 \leq 0.2$) lie partially
  above the heavy black line, indicating that growth in this region is
  spontaneous, as in Figure \ref{fig_2}. In Griffith's analysis, load is held
  constant ($p/p_G = 1$) and spontaneous growth is unabated. 
  This condition is represented by the horizontal line (abscissa $p/p_G=1$) where
  the line $\alpha_0 = 0$ would plot.
  \label{fig_3}}
\end{figure}

Crack growth for systems where $\alpha_0 > 0.2$ and for fracture
growth beyond the region of spontaneous growth in systems where
$\alpha_0 < 0.2$ is made possible by injecting fluid into the cavity
to raise the pressure.  The rate at which fluid must be injected to
maintain fracture growth is found by first considering the changes in
fluid pressure during growth shown in Figure \ref{fig_3}. The fluid
pressure $p_f$ required for continued growth is found by equating
(\ref{eqn17}) and (\ref{eqn18}) giving:

\begin{equation}
p_f/ p_G = \left( c / c_0 \right)^{-1/2}
\label{eqn24}
\end{equation}

Pressure $p_f/p_G$ in (\ref{eqn24}) is represented graphically by the
heavy black line in Figure \ref{fig_3}. The thin lines in the figure give the
pressure from (\ref{eqn20}) in the cavity if the fracture were to grow
with no fluid injection.

Consider now the growth of the fracture after injection has brought
fluid pressure $p$ to the Griffith value $p_G$ in (\ref{eqn14}).  A
differential increase $dc$ in fracture radius with no injection
decreases fluid pressure by $dp/p_G$ in Figure \ref{fig_3}; that is,
from (\ref{eqn20}),

\begin{equation}
  \frac{dp}{p_G} = \frac{d}{dc} \left( \frac{1+\alpha_0}{1+\alpha_0 \left(c/c_0 \right)^3} \right) dc.
\label{eqn25}
\end{equation}

Increasing fracture radius causes a differential change $dp_f/p_G$
in the non-dimensional fluid pressure required for continued growth. We find from
(\ref{eqn24}) that the change $dp_f/p_G$ is:

\begin{equation}
\frac{dp_f}{p_G} = \frac{d}{dc} \left( c/c_0 \right) ^{-1/2} dc.
\label{eqn26}
\end{equation}

For continued growth of the fracture, fluid pressure $p_f$ in
(\ref{eqn24}) must be maintained, i.e., the rate of pressure change
$\left( dp_f - dp \right) / p_G$ must be compensated for by
injecting fluid into the crack. We see from (\ref{eqn4}), that the
rate $dV_i$ at which fluid must be injected for a fracture having
radius $c/c_0$ is given by the expression:

\begin{equation}
  d V_i = \beta_f V_f d\left(p_f - p \right) \left[ 1 + \alpha_0 \left( c/c_0 \right)^3 \right],
\label{eqn27}
\end{equation}

\noindent or, in non-dimensional form:

\begin{equation}
  \frac{dV_i}{V_G} = \frac{d\left( p_f - p \right)}{p_G} \left[ \frac{1 + \alpha_0 \left( c/c_0 \right)^3}{1 + \alpha_0} \right],
\label{eqn28}
\end{equation}

\noindent
where $V_G$ is the volume of fluid necessary to force growth of the
initial crack ($c = c_0$) given by:

\begin{equation}
V_G = \beta_f V_f p_G.
\label{eqn29}
\end{equation}

Combining (\ref{eqn24})-(\ref{eqn28}), one finds that the rate $\left(
  dV_i / dp \right)_f$ at which fluid must be injected to maintain
growth per unit change in fluid pressure is given by the expression:

\begin{equation}
  \left[ \frac{d \left( V_i / V_G \right)}{ d \left( p / p_G \right)}\right]_f = \frac{1-5 \alpha_0 \left( c/c_0 \right)^3}{1 + \alpha_0}.
\label{eqn30} 
\end{equation}

\begin{figure}
\centering
\includegraphics[width=0.6\textwidth]{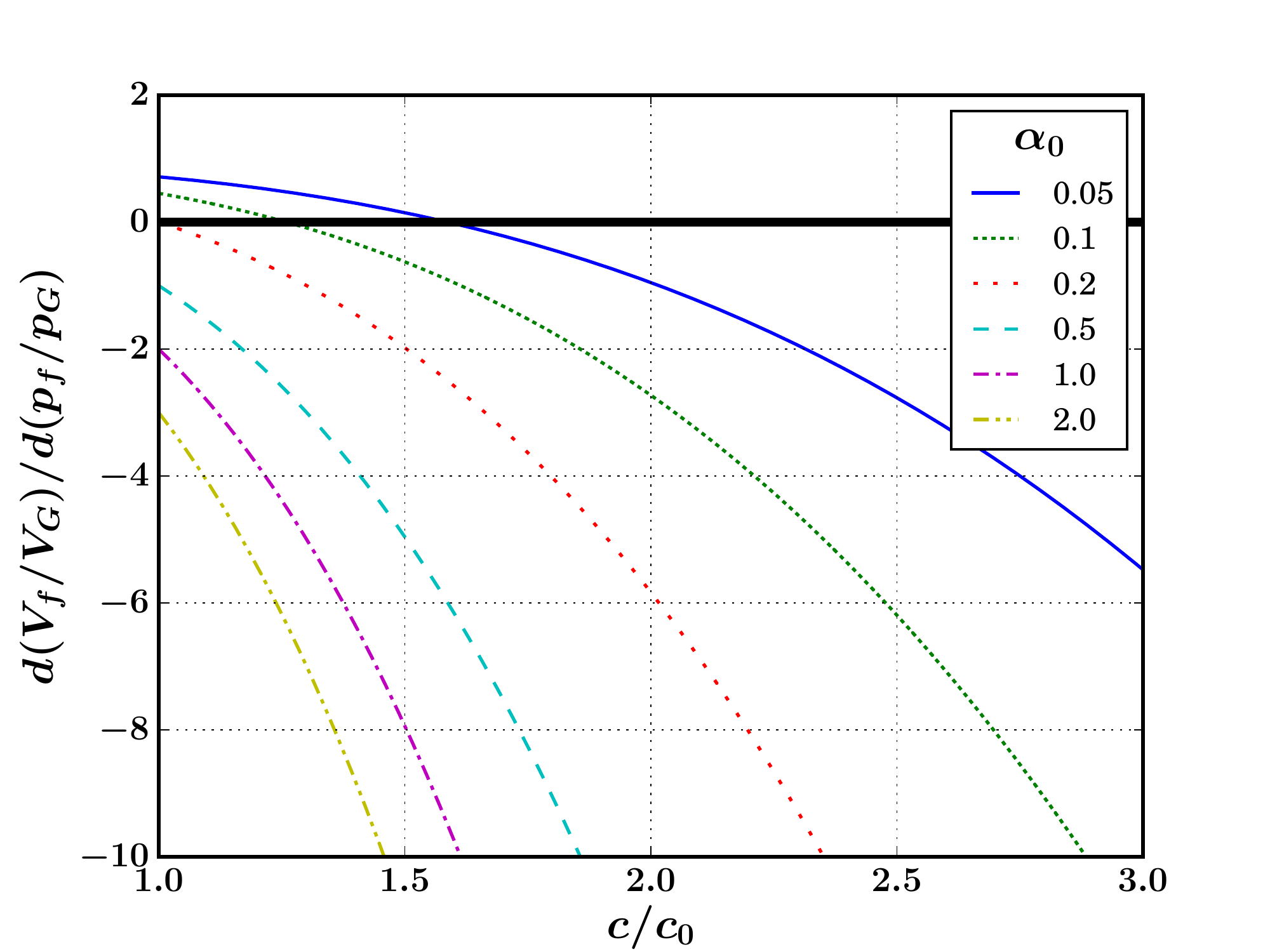}
\caption{We show in Figure \ref{fig_3} the increase in fluid pressure $p_f$
  needed to maintain growth once the fracture has been initiated at
  $p_G$. Here we plot the volumetric rate at which fluid must be injected per
  unit change in fracture pressure to maintain growth as given by
  (\ref{eqn4}) and(\ref{eqn30}). Note that the injection volume rate
  depends strongly on $\alpha_0$. The heavy black line denotes the reference
  level where no fluid is injected after fracture
  initiation. \label{fig_4}}
\end{figure}

\noindent
The rate at which fluid must be injected per unit change in fracture
pressure to maintain growth given by (\ref{eqn30}) is plotted in Figure
\ref{fig_4}.

\section{Discussion}

In our analysis, we find that the pressure required to initiate
hydraulic fracturing is the same as that derived by
\citet{griffith:1921,griffith:1924}.  The subsequent growth of the
crack depends on the parameter $\alpha_0$ (see (\ref{eqn4}) and
(\ref{eqn5})), where $\alpha_0$ is the ratio of the compliance of the
cavity in the country rock surrounding the fracture relative to the
compliance of the fracturing fluid itself. The two phases, the cavity
and the fluid, include the entire pressurized volume, i.e., both the
fracture and the borehole between the packers. We assume, without loss
of generality, that the system is in equilibrium and fluid pressure is
zero when fluid injection is initiated.

The parameter $\alpha_0$ arises in the first step in the analysis,
namely our derivation of the increase $dp$ in pressure in the cavity
resulting from injecting a differential volume $dV_i$ at `zero'
(i.e. datum) pressure into the cavity. We follow Eshelby's
\citep{eshelby:1957} technique for calculating the changes in
stress and strain arising from an inhomogeneity undergoing a
transformational strain; the procedure is outlined in the text leading
to the final expression (\ref{eqn4}).

One can see in equation (\ref{eqn5}) definition of the ratio of the
compliances of the fluid and cavity phases. The compliance of the
fluid component is simply $\beta_f V_f$ where $\beta_f$ is the
compressibility of the fluid and $V_f$ is the total volume of the
fluid. The fluid is under the same hydrostatic
pressure everywhere (for the equilibrium conditions we assume) and so
the compliance $\beta_f V_f$ is exact.

On the other hand, the term defining the compliance of the cavity in
(\ref{eqn5}) is exact only for a penny-shaped crack, whereas the
deformation of the cavity under pressure includes that of the borehole
and the packers. This simplification is acceptable, in our opinion. In
the first place, deformation of the hydrofracture is likely to
dominate the overall deformation of the cavity because it is the most
compliant component -- the relatively rigid borehole/packer system
contributes to the volume $V_f$ but has relatively little influence on
compliance. Also, we are primarily concerned with changes in crack
radius, which are not coupled to the borehole/packer unit. And
finally, the goal of this analysis is only to point out the important
role of $\alpha_0$ in the mechanics of hydrofracturing, and a simple
model like ours is not only adequate, but also desirable.

To calculate the fluid pressure $p_{G}$ required to initiate fracture,
we follow Griffith's \citep{griffith:1921,griffith:1924} procedure,
with one additional step. Griffith, in his analysis, increased the
stress acting on a  crack until the work done under constant load as the
fracture advanced just equaled the energy needed to create new
fracture surface. In our model, increasing the fracture size increases
the volume of the cavity causing the fluid pressure to drop.  
Analysis of the components of work performed during a
  differential increase in fracture radius (equations (\ref{eqn6})
  through (\ref{eqn14}))
shows that pressure $p_f$ required to
initiate fracture is Griffith's value $p_G$ even though pressure
decreases as the fracture starts to grow.

Although fracture pressure $p_G$ is a necessary condition for growth
of the hydrofracture, we find in the section \textit{Fracture Growth}
that it is sufficient over only a range of values of $\alpha_0$. We
analyze the advance of the fracture by stipulating that the Griffith
criterion must be met at each stage of growth. Our results are
summarized in Figures \ref{fig_2}, \ref{fig_3}, and \ref{fig_4}. In
Figure \ref{fig_2}, we see Griffith's criterion in graphical
form. Griffith postulated that the strain energy release rate must be
equal to or greater than the rate at which surface energy is
dissipated. We see that this condition is met at $c/c_0 = 1.0$ for any
value of $\alpha_0$. Note, however that, as the fracture grows,
Griffith's criterion is met only over a range of values of $\alpha_0$
($0 \leq \alpha_0 \leq 0.2$). Over this range the fracture radius
increases spontaneously and then stops. In cases where $\alpha_0 >
0.2$ or where $\alpha_0 < 0.2$ but beyond the range of spontaneous
growth, fracture size can be increased only by increasing energy by
increasing fluid pressure.

In Figure \ref{fig_3} we have plotted the increase in fluid pressure needed to
increase fracture radius. The heavy black line defines the pressure $p_f$
needed for growth and the thin lines give the pressure $p$ in the
fracture when fluid volume remains constant after initiation. We see
in the figure that fluid pressure is always sufficient for growth for
systems where $\alpha_0$ is in the range ($0 \leq \alpha_0 \leq
0.2$). This, of course is expected from Figure \ref{fig_2} where fractures are
seen to grow spontaneously when $\alpha_0$ is in this range. Fluid
must be injected for growth when $\alpha_0 > 0.2$ or to activate
fractures which have stopped after growing spontaneously.

The rate $\left( dV_i / dp \right)_f$ at which fluid must be injected
per unit change in pressure to maintain fracture growth is calculated
(see (\ref{eqn27})) and plotted in Figure \ref{fig_4}. Note in the figure that
the rate is uniform and positive while the initial crack ($c=c_0$) is
pressurized. The rate changes immediately when fluid pressure reaches
Griffith fracture pressure, $p_G$ and the fracture begins to grow; the
change depending on the value of $\alpha_0$ for the system. We see in
the figure for systems where $\alpha_0$ is in the range $0 \leq
\alpha_0 \leq 0.2$ that the rate $\left( dV_i / dp \right)_f$ remains
positive, indicating that fluid is returning to the well head
$\left( dV_i < 0 \right)_f$ while pressure is decreasing $\left( dp_f < 0
\right)$. In those regions in the figure where negative values of
$\left( dV_i / dp \right)_f$ are found, fluid pressure is decreasing
while the fracture is growing, even though fluid is being injected
into the fracture.

One parameter that we have ignored in the analysis is time. We have
assumed that fluid completely fills the fracture during deformation
and fluid pressure is the same throughout the cavity. This assumption
is questionable for operations involving viscous fluids, fast growing
fractures, fractures with very small apertures, etc. In an attempt to
get an idea of how much time might affect our results, we analyzed an
extreme case where the fluid and fracture properties are such that the
fluid remains in the initial cavity while the hydrofracture
advances. 

The model is that in Figure \ref{fig_1} except that the load is
provided by fluid pressure applied over only the area of the initial
defect ($c/c_0 \leq 1.0$), and the fracture remains `dry' as it
advances.  The analysis follows the same steps as those described
above, except that elastic strain energy is derived following standard
procedure from the stress concentration factor found by Sneddon
\citep{tada:2000}.  Details of
the calculation are superfluous in a study like the one here; suffice
to say, all steps in the procedure outlined in the text leading to
Figures \ref{fig_2}, \ref{fig_3}, and \ref{fig_4} are the same,
requiring only a redefinition of radius $c/c_0$. We find that the
elastic strain energy release rate drops precipitously once Griffith's
fracture initiation pressure $p_G$ is reached, and fractures do not
advance spontaneously for any values of $\alpha_0$. Further, the
pressure increase, or the fluid volume injected, required to make the
fracture grow is much more than needed for the `wet' case shown in
Figures \ref{fig_3} and \ref{fig_4}. These results suggest that it is
energetically favorable for the fracture to grow at the rate
prescribed by the analysis of the `wet' case described in the text
where equilibrium conditions prevail. To carry the calculation further
requires analyzing the interrelationship between the flow of fluid and
the rate at which the crack advances, an interesting problem but one
that is far beyond the scope of a preliminary study such as the one
here.

\newpage
\section{ACKNOWLEDGMENTS}
We thank Bezalel Haimson, John Rudnicki, Nafi Toksoz, and Michael
Fehler for valuable discussions.  Our thanks also go to Brice
Lecampion and John Rudnicki who pointed out a significant error in an
earlier version of the analysis and to Jared Atkinson for a critical
review.

\vspace{\baselineskip}
\noindent
\textit{\textbf{Note from coauthor Stephen Brown:} I worked with Joe
  Walsh off and on throughout my entire career. It was a pleasure to
  have worked closely with him from 2013-2015 during the last few
  years of his life on this paper considering the physics of hydraulic
  fracturing. Unfortunately, it was never published. I am pleased 
  to be able to now provide this work to the science and engineering
  community as an open access reference for educational use. This
  paper is likely Joe's final research contribution. We are all
  grateful for him. Rest in peace, Joe.}

\vspace{\baselineskip}
\noindent
\textit{\textbf{Please read more about Joe Walsh's life here:} Scholz, C. H.,
  Goldsby, D. L., Bernab\'{e}, Y., and Evans, B., (2018), Joseph
  B. Walsh (1930–2017), Eos, \textbf{99},
  https://doi.org/10.1029/2018EO093999. Published on 6 March 2018.}

\vspace{2\baselineskip}



\begin{thebibliography}{99}

\bibitem[\textit{Eshelby}, 1957]{eshelby:1957} Eshelby, J. D., (1957),
  The determination of the elastic field of an ellipsoidal inclusion,
  and related problems, Proceedings of the Royal Society of London A,
  \textbf{241}, 376-396.

\bibitem[\textit{Griffith}, 1921]{griffith:1921} Griffith, A. A.,
  (1921), The phenomena of rupture and flow in solids, Philosophical
  Transactions of the Royal Society of London, \textbf{A221}, 163-198.

\bibitem[\textit{Griffith}, 1924]{griffith:1924} Griffith, A. A.,
  (1924), The theory of rupture, \textit{in} Proceedings of the First
  International Congress on Applied Mechanics, C.B. Biezeno and
  J.M. Burgers, eds., J. Waltman Jr, Delft, 55-63.

\bibitem[\textit{Love}, 1944]{love:1944} Love, A. E. H., (1944), A
  treatise on the mathematical theory of elasticity (4th ed). New York
  Dover Publications.

\bibitem[\textit{Sneddon}, 1946]{sneddon:1946} Sneddon, I.N., (1946),
  The distribution of stress in the neighborhood of a crack in an
  elastic solid, Proceedings of the Royal Society of London A22,
  \textbf{187}, 229-260.

\bibitem[\textit{Sack}, 1946]{sack:1946} Sack, R. A., (1946), Extension
  of Griffith's theory of rupture to three dimensions, Proceedings of
  the Physical Society, \textbf{58}, 729--736.


\bibitem[\textit{Tada, et al.}, 2000]{tada:2000} Tada, H.,
  P.P.C. Paris, and G.R. Irwin, (2000), The Stress Analysis of Cracks
    Handbook, American Society of Mechanical Engineers, 696 p.

\end{thebibliography}
\end{document}